\documentclass[fleqn,11pt]{SelfArx} 

\usepackage[english]{babel} 

\usepackage[locale=UK]{siunitx}
\usepackage{graphicx}
\usepackage[numbers]{natbib}
\usepackage{subcaption}
\usepackage[abs]{overpic}
\usepackage[toc,page]{appendix}
\usepackage{braket}

\definecolor{color1}{RGB}{0,0,90} 
\definecolor{color2}{RGB}{0,20,20} 

\usepackage{hyperref} 
\hypersetup{hidelinks,colorlinks,breaklinks=true,urlcolor=color2,citecolor=color1,linkcolor=color1,bookmarksopen=false,pdftitle={Title},pdfauthor={Author}}

\JournalInfo{ArXiv, Mar 2017} 
\Archive{} 

\PaperTitle{Room temperature bulk diamond \mbox{\textsuperscript{13}C hyperpolarisation} -- Strong evidence for a complex four spin coupling}

\Authors{Ralf Wunderlich\textsuperscript{1}*, Jonas Kohlrautz\textsuperscript{1}, Bernd Abel\textsuperscript{2}, Jürgen Haase\textsuperscript{1} and Jan Meijer\textsuperscript{1}} 
\affiliation{\textsuperscript{1}\textit{Faculty of Physics and Earth Sciences, Felix Bloch Institute for Solid State Physics, Leipzig University, Linnéstrasse 5, 04103 Leipzig, Germany}} 
\affiliation{\textsuperscript{2}\textit{Leibniz-Institute of Surface Modification (IOM), Permoserstrasse 15, 04318 Leipzig, Germany}} 
\affiliation{*\textbf{Corresponding author}: ralf.wunderlich@uni-leipzig.de} 

\newcommand{\keywordname}{Keywords} 

\Abstract{
Hyperpolarisation at room temperature is one of the most important research fields in order to improve liquid, gas or nanoparticle tracer  for Magnetic Resonance Imaging (MRI) in medical applications.
In this paper we utilize nuclear magnetic resonance (NMR) to investigate the hyperpolarisation effect of negatively charged nitrogen vacancy (NV) centres on  carbon-13 nuclei  and their spin diffusion in a diamond single crystal close to the excited state level anti crossing (ESLAC) around \SI{50}{\milli\tesla}.
Whereas the electron spins of the NV centre can be easily polarized in its m = 0 ground state at room temperature just by irradiation with green light , the swop of the NV electron spin polarization to a carbon-13 nuclei is a complex task.
We found that  the coupling between the polarized NV electron spin, the electron spin of a substitutional nitrogen impurity (P1) as well as its nitrogen-14 nuclei and the carbon-13 nuclear spin has to be considered.
Here we show that through an optimization of this procedure, in about two minutes a signal to noise ratio which corresponds to a 23 hour standard measurement without hyperpolarisation and an accumulation of 460 single scans can be obtained.
Furthermore we were able to identify several polarisation peaks of different sign at different magnetic fields in a region of some tens of gauss. Most of the peaks can be attributed to a coupling of the NV centres to nearby P1 centres.
We present a new theoretical model in a framework of cross polarisation of a four spin dynamic model in good agreement with our experimental data. The results demonstrate the opportunities and power as well as limitations of hyperpolarisation in diamond via NV centres. We expect that the current work may have a significant impact on future applications.
}

\begin{document}

\flushbottom 

\maketitle 

\tableofcontents 

\thispagestyle{empty} 


\section*{Introduction} 

\addcontentsline{toc}{section}{Introduction} 
Magnetic resonance tomography (MRT) is one of the most important tools of today's modern medicine diagnostics.
Nuclear Magnetic Resonance (NMR) in general benefits from high spectral resolution but suffers of low sensitivity which is in a $1/2$ spin system proportional to the spin polarisation $P=\left[ N_{\text{u}} - N_{\text{d}} \right] / \left[ N_{\text{u}} + N_{\text{d}} \right]$ and the number of spins $N=N_{\text{u}} + N_{\text{d}}$. Due to the Boltzmann distribution this fraction is just \num{e-5} at room temperature and standard conditions in humane medicine.
An enhancement of this factor is directly related to the lateral resolution and could increase the quality of MRT by orders of magnitude.
MRT is based mainly on the visualization and the dynamics of proton spins. However in the last century a large number of new tracers have been tested in order to visualize dynamic processes inside the body. Functionality tests using hyperpolarised \textsuperscript{3}He or \textsuperscript{129}Xe nuclei in the gas phase for lung cancer is an example and routinely applied nowadays but still under research and development \cite{Mugler2013}.
In contrast the electron spin of a negativly charged nitrogen vacancy (NV) centre, consisting of a substitutional nitrogen atom next to a vacancy in the diamond lattice, can be nearly polarized to \SI{100}{\percent} by irradiation  with green light \cite{Doherty2013}.
Common goal of a number of research groups is the transfer of the electron polarisation to nuclear spins. 
The NV centre - carbon-13 nucleus structure is an outstanding model system which combines the possibility of a nearly fully polarised electron spin reservoir and a easy to measure nuclear spin species.
In thermal equilibrium the polarisation is given by $P=\tanh{\left({\hbar \omega}/{k_{\text{B}} T}\right)}$ and a brute force method to enhance the signal is cooling down the spin system. This can be combined with the optical induced polarisation of the the NV centre in high magnetic fields which also leads to a non-equilibrium spin polarisation of C-13 nuclei in diamond \cite{King2010,Scott2016}. Also more sophisticated microwave driven nuclear polarisation (DNP) techniques in the manner of Overhauser \cite{Overhauser1952} like the solid effect (SE) \cite{HovavSE2010} the cross effect (CE) \cite{HovavCE2012} or the thermal mixing \cite{Hu2011} can be used and are already combined with the properties of NV centres \cite{Yang2016,Scheuer2016, Alvarez2015}.
An approach without additional micro wave appliancation to overcome the low signal to noise ratio (SNR) is to deploy the hyperfine or dipole interaction between a polarised and an unpolarised spin system. This was done in Ref. \cite{Fischer2013,Fischer2013a} in which the authors attribute their polarisation effect to the exited state level anti-crossing (ESLAC), which takes place at around \SI{51}{mT}.
Previous optical hole-bleaching experimenal studies reported already cross polarisation of negatively charged NV and substitutional nitrogen (P1) centres by field dependent absorption messurements with a narrow-band laser \cite{Holliday1989}. However this technique could not distingush between up or down polarisation and disregarding nuclear spins. More recently this effect was published for single NV centres as well as for neutral NV centres detected by Optical Detected Magnetic Resonance (ODMR) \cite{Hanson2006,Belthangady2013,Wang2014}. Both consider only coupling between the electron spins.
To exploit the nuclear hyperpolarisation effect we use a modified NMR system. With this approach we are sensitive to carbon-13 spins including those far away from all ODMR-active centres and other defects.


\section{Methods}
\begin{figure}[tb]
\centering
\includegraphics[width=\linewidth]{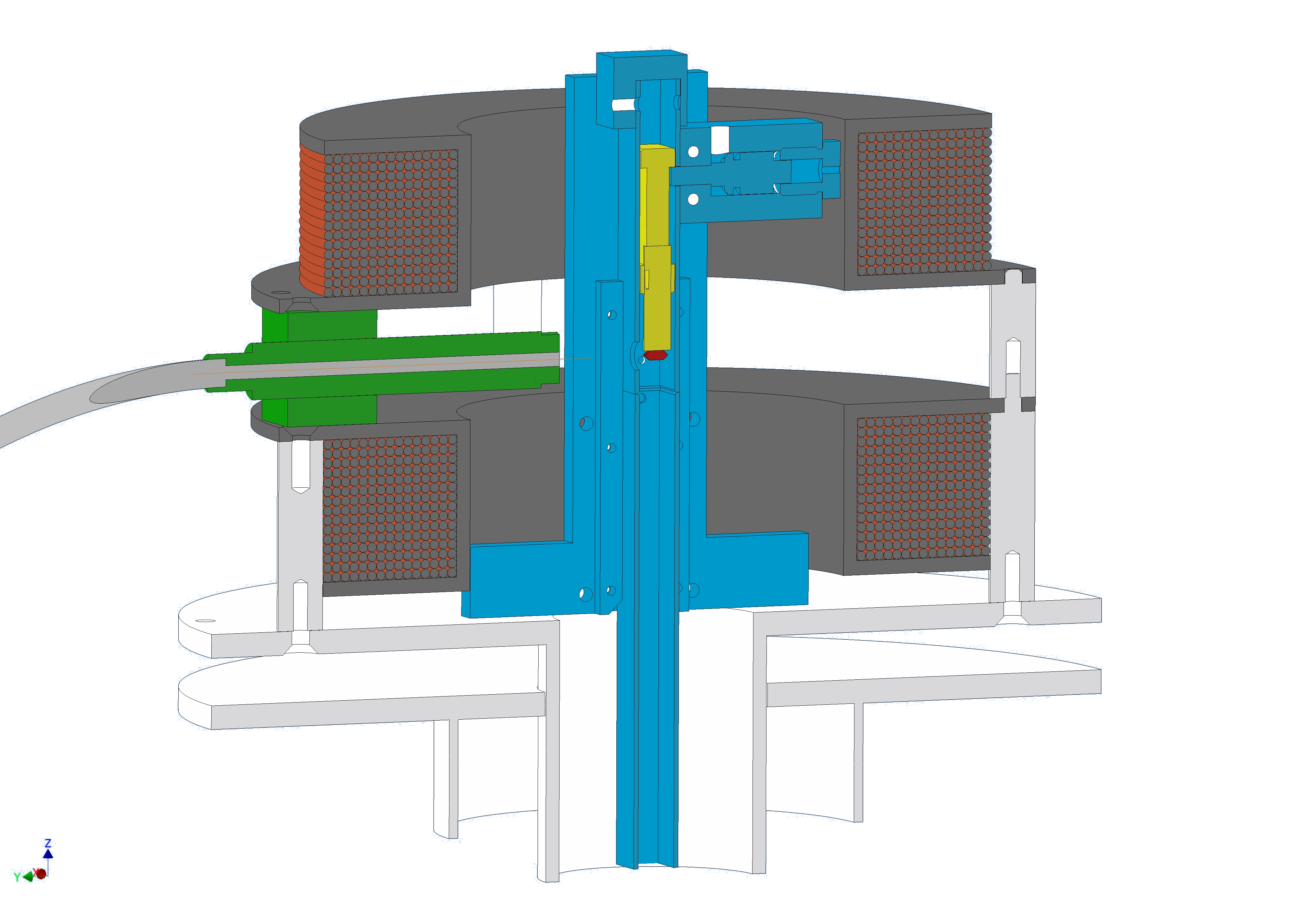}\llap{\includegraphics[height=2cm]{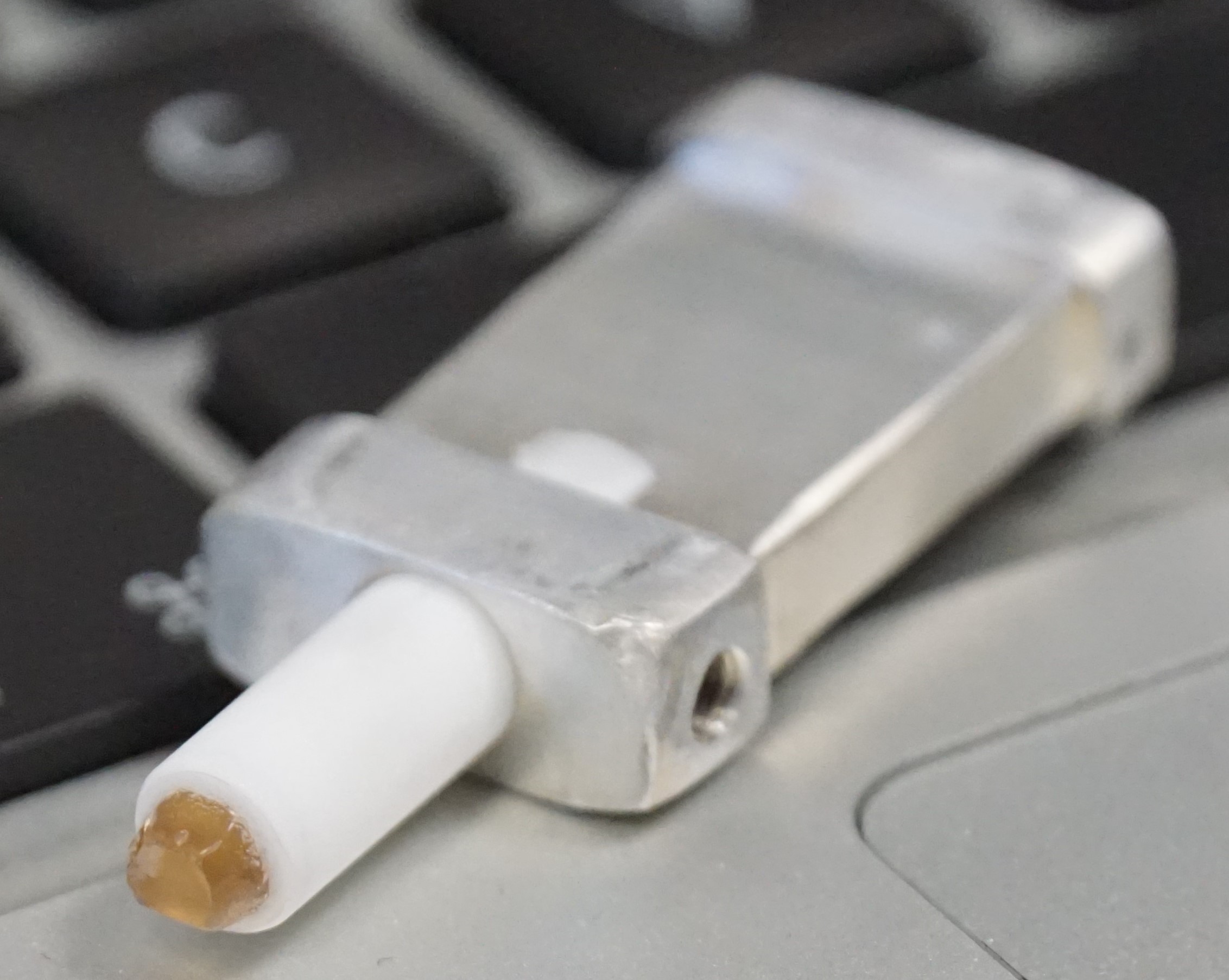}}
\caption{Sketch of the low field irradiation unit. The sample (red) is mounted on the shuttle (yellow) and fixed in the pneumatic shuttling system (blue). The Helmholtz configuration of the coils (grey) in superposition with the stray field of the NMR magnet allows to apply easily a field up to \SI{60}{\milli\tesla}. Between the coils one has an optical access for the laser irradiation (green). The inset shows a picture of the sample on top of the shuttle.}
\label{lowFieldUnit}
\end{figure}

A home-made Helmholtz coil in combination with a programmable power supply provides a constant low magnetic field with a stability of about \SI{0.1}{\milli\tesla}.
The diamond sample is fixed on a metal shuttle and aligned with one of the four NV axes along the magnetic field vector.
For the NV polarisation we use a \SI{532}{nm} laser with a power up to \SI{5}{\watt}. The laser light is guided through a \SI{400}{\micro \meter} core multimode optical fiber with an numerical aperture of \SI{0.22}{}. The SMA905 connector at the end of the fiber is screwed several millimeters before the sample (Fig. \ref{lowFieldUnit}).
The shuttle is guided inside a rectangular hollow profile from the low field coils to a probehead in the middle of a commercial superconducting \SI{300}{\mega \hertz} NMR magnet. After a shuttling duration of about one second the sample is placed between a pair of Helmholtz-like RF coils with a diameter of \SI{6}{\milli\meter} fixed on a ceramic frame. Subsequently a $\pi /2$ pulse of \SI{15}{\micro s} is applied and the NMR signal of the C-13 spins is recorded.
Afterwards the shuttle is lifted up via a pneumatic system back into the radiation unit and is fixed. After short cooling by several air jets the next run with different settings e.g. magnetic field or laser duration starts completely automatically.
This enables a systematic and reproducible investigation of the hyperpolarisation process.
In comparison to ODMR techniques the presented approach allows the observation of the polarisation effect of nuclei far away from the ODMR active centre. This is in particular important to investigate spin diffusion processes in solids and consequently a development of an overall bulk polarisation and application.\\
All measurements were performed with a commercially available HPHT diamond sample from Element Six with dimensions of about {$2 \times 2 \times 2$ mm$^3$} and a naturally abundance of \SI{1.1}{\percent} C-13 isotopes moreover a nitrogen content of \numrange{100}{200} ppm. After a \SI{10}{MeV} electron irradiation with a fluence of \SI{2 e 16 }{\per\centi\meter\squared} the sample was annealed at \SI{800}{\degreeCelsius} for \SI{2}{\hour} (tab.\ref{properties}).
\begin{table}[tb]
\caption{Sample properties}
\centering
\begin{tabular}{llr}
\toprule
\multicolumn{2}{c}{ID 179 (111)-diamond} \\
\cmidrule(r){1-2}
property & value & unit \\
\midrule
mass & 0.0738 & \si{\gram} \\
N content & 100 - 200 & ppm \\
C-13 content & \SI{1.1}{} & \si{\percent} \\
e\textsuperscript{-} irradiation fluence & \num{2 e 16} & \si{\per\centi\meter\squared} \\
annealing temperature & \num{800} & \si{\degreeCelsius} \\
annealing time & \num{2} & \si{\hour} \\
\bottomrule
\end{tabular}
\label{properties}
\end{table}
During irradiation the sample reaches a maximal temperature of about \SI{190}{\degreeCelsius}. Before and after each preparation step an photoluminescence  spectrum was taken during a {$80 \times$\SI{80}{\micro m}} x-y-Scan \SI{50}{\micro m} below the sample surface in a self-made confocal microscope with an air objective MPlanoApo \SI{100}{x}/0.95 from Olympus to monitor the NV creation and its density.
For excitation we use a \SI{532}{\nano \meter} Nd:YAG Laser and the spectrometer containing a \SI{150}{\per\milli\meter} grid as monochromator from Jobin Yvon with a CCD sensor of \SI{2048}{pixels}.


\section{Results and Discussion}
\subsection{Optical Spectrum}
\begin{figure}[tb]\centering
\includegraphics[trim = 20mm 10mm 30mm 10mm, width=\linewidth]{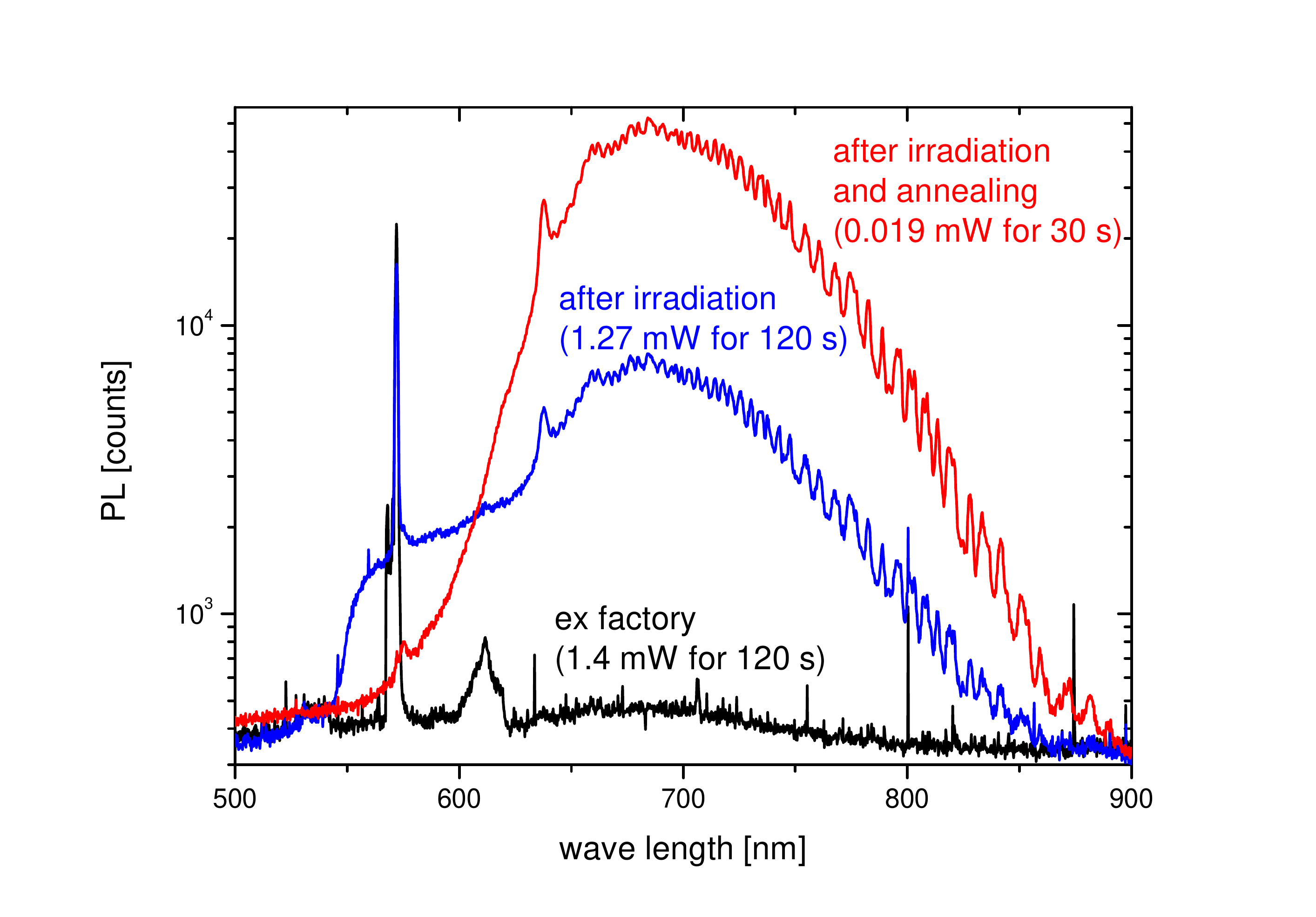} 
\caption{Optical spectra of photo luminescence (PL) before and after the single sample preparation steps. The laser power and acquisition times for each spectrum are indicated in diagram. Note the smaller excitation power and acquisition time for the spectrum after irradiation and annealing. The small oscillations between \SI{700}{\nano\meter} and \SI{850}{\nano\meter} above the NV peak at \SI{637}{\nano\meter} arise from interference inside the spectrometer.}
\label{OptSpecComp}
\end{figure}

Figure \ref{OptSpecComp} shows three different photo luminescence (PL) spectra each after specific preparation step. For the original sample ex factory (black) just the Raman peak at \SI{573}{\nano \meter}, a second feature at about \SI{612}{\nano\meter} and no NV signal is noticeable. The \SI{612}{\nano\meter} peak is actually known as PL from some natural type I diamonds \cite{Zaitsev2001}.
After electron irradiation a clear $\text{NV}^-$ zero phonon line (ZPL) at \SI{637}{\nano\meter} with its characteristic side bands arises and the \SI{612}{\nano\meter} peak disappears (blue). Besides the Raman peak a small $\text{NV}^0$ at \SI{575}{\nano\meter} was vaguely perceptible.
Subsequent annealing leads to an massive boost in the NV signal (red).
Please note while the spectrum for the untreated (black) and electron irradiated (blue) sample is recorded under similar conditions apparent from similar heights of the Raman peaks the spectrum after annealing (red) is recorded four times shorter and with only \SI{1.5}{\percent} of the laser power. This corresponds to a factor of $1400$ in signal intensity and therefore NV density compared to the sample that was untreated and not annealed.

\subsection{NMR measurements}
\begin{figure*}[tb]\centering 
\centering
	\begin{minipage}{0.70\linewidth}
        \centering
        \includegraphics[width=\linewidth]{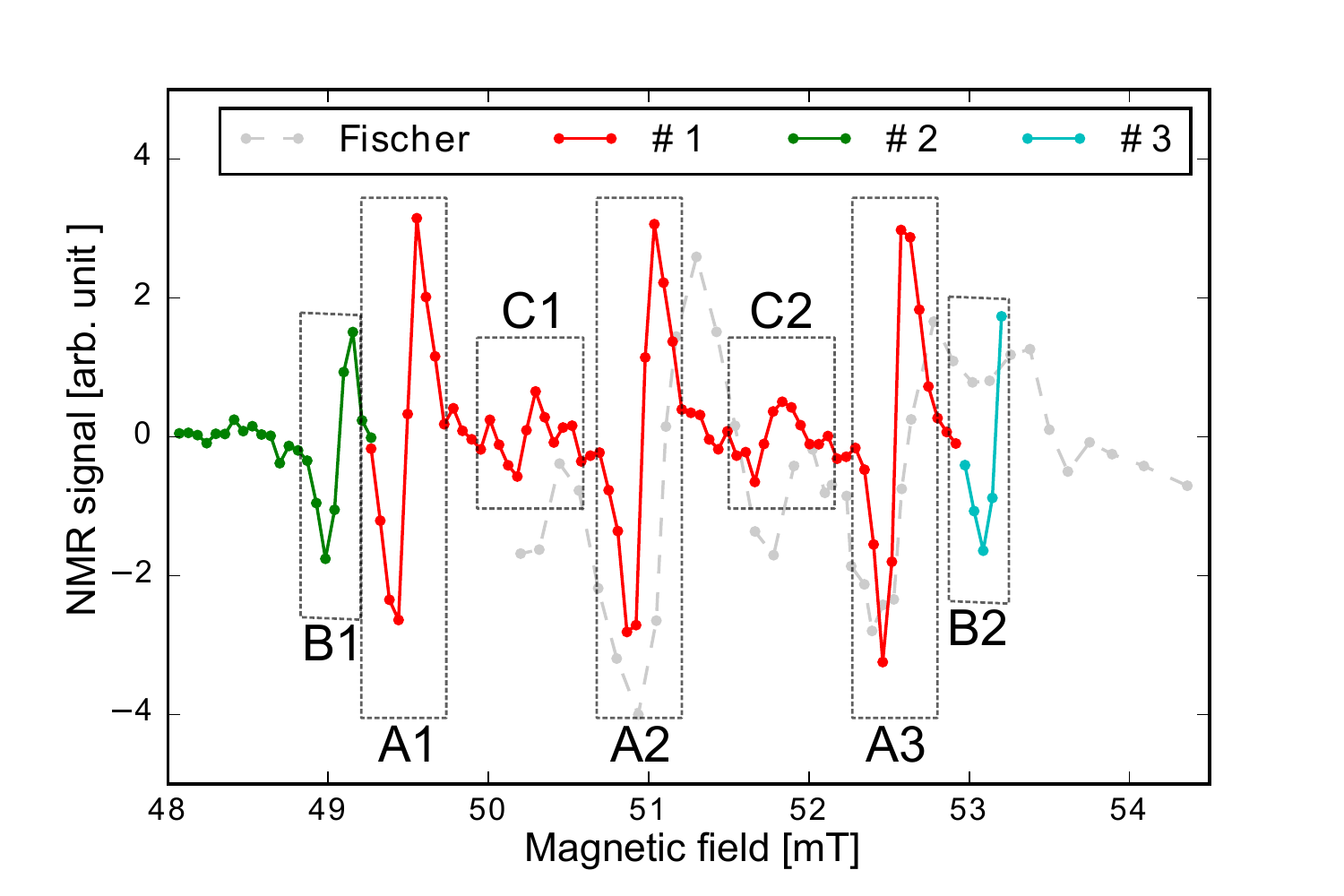}
        \put(-340,202){a)}
    \end{minipage}
	\begin{minipage}{0.25\linewidth}
        \includegraphics[width=\linewidth]{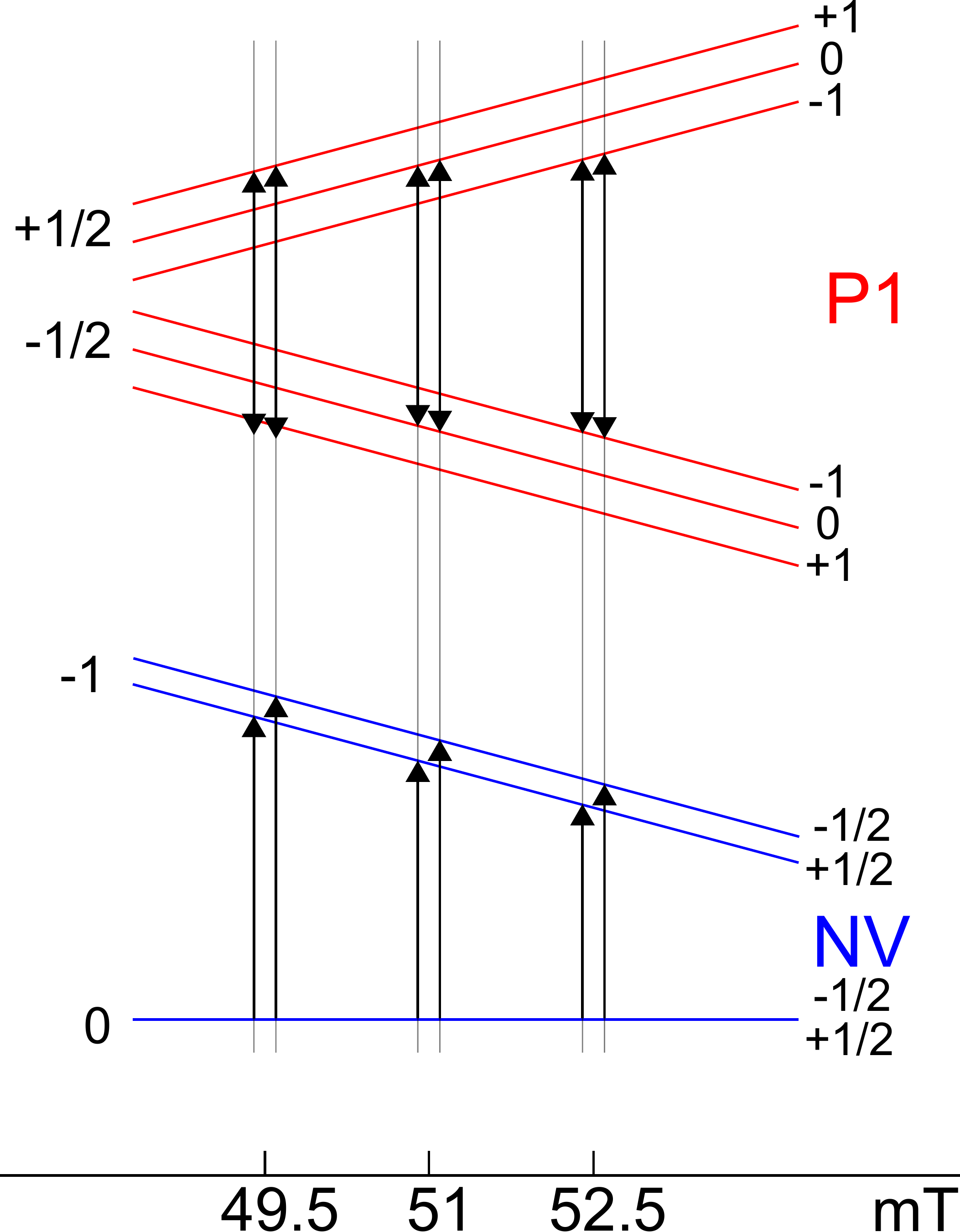}
        \put(-140,180){b)}
    \end{minipage}
\caption{Integral NMR signal intensity after \SI{120}{\second} irradiation with a green \SI{5}{\watt} laser at varying magnetic fields (a). The various colors indicate different data sets - however acquired under similar conditions. Note the signal is after one single shot, not accumulated and scaled to compare with the data from Fischer et al \cite{Fischer2013}. The sketch in b) depicts the energy level diagram and the observed transitions.}
\label{fieldepFull}
\end{figure*}

All presented measurements were conducted at ambient conditions and the noise was checked to be thermal.
First we estimate the $\pi/2$ pulse length with a nutation experiment at about \SI{7}{T} to \SI{15}{\micro\second} with a carrier frequency of \SI{75.4689}{\mega\hertz}.
Figure \ref{fieldepFull} depicts the integral NMR signal intensity depending on the magnetic field during laser irradiation and a shuttling into the NMR probe within about one second.
\begin{figure}[tb]
\begin{center}
\begin{overpic}[width=\linewidth]{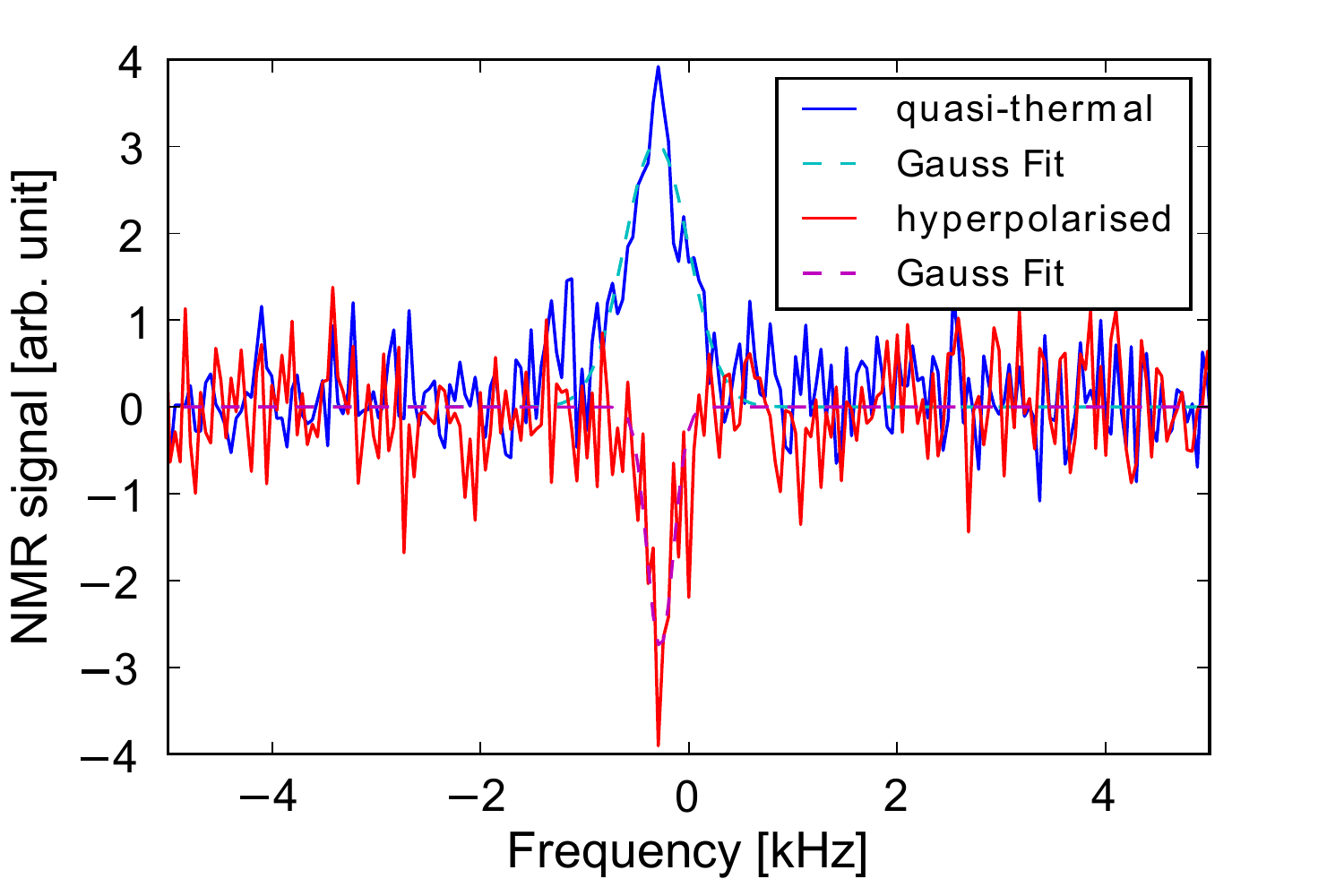}
\put(38,27){\includegraphics[trim=0 0 0 0, clip, height=1.75cm]{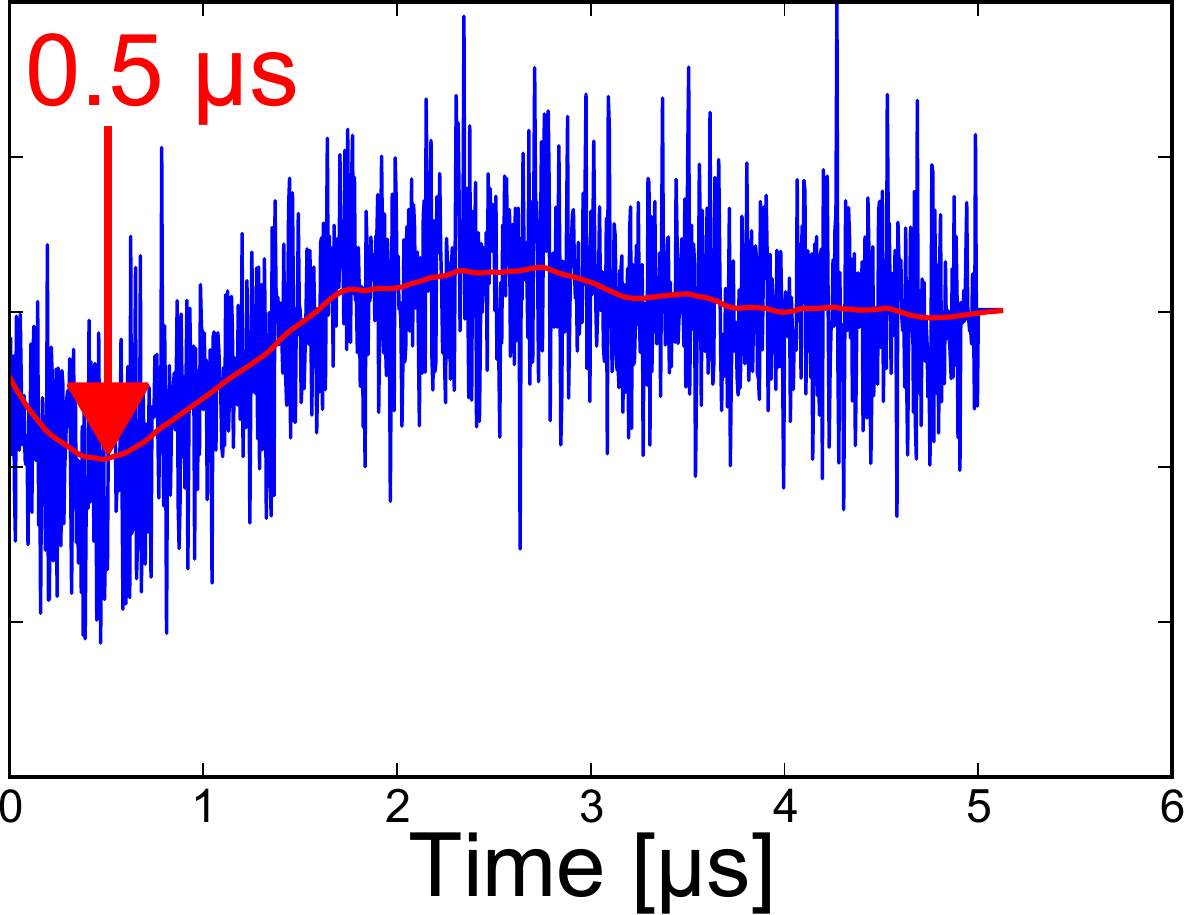}}
\end{overpic}
\caption{Comparison of NMR spectra accumulated in quasi-thermal equilibrium (blue) over 267 scans  separated by \SI{640}{\second} (over all \SI{47.5}{\hour} ac\-cu\-mu\-la\-ion time) and a single scan after hyper-\-polarisation during \SI{400}{\second} irradiation at \SI{49.35}{\milli\tesla} (red). In the former case no signal can be found after a single scan. The quasi-thermal spectrum is scaled in the intensity for a better comparison. Note the opposite sign but yet similar noise level for both data sets. The inset shows the signal of a solid echo experiment with $\tau=\SI{500}{\micro\second}$ in the time domain.}
\label{spectra}
\end{center}
\end{figure}
One representative spectrum of the acquired spectra and the corresponding thermal spectrum are plotted in Fig. \ref{spectra}. Both spectra were fitted to a Gaussian function. The thermal spectrum (a) has a linewidth (FWHM) of \SI{733\pm 22}{\hertz} and the hyper-polarised one has a smaller width of \SI{298\pm 12}{\hertz}, both based on the fits. However at least the hyperpolarised spectrum has an ever narrower linewidth slightly below this value regarding the raw data.
This discrepancy can be explained with the rather short last delay of \SI{640}{\second} due to the requirement of an adequate accumulation time in the thermal case compared to the measured spin-lattice relaxation time of some hours (see below). This means that in the thermal experiment we measure the C-13 nuclei with shorter T1 times in proximity to impurities like P1 centres which are indeed present all over the investigated sample (see nitrogen content in Tab.\ref{properties}).
The narrow line width in the hyper-polarised case can be compared with theoretical calculation of the line width of C-13 nuclei solely dipolar coupled to each other. An estimation for homonuclear dipolar coupling by numerical simulations lead to a linewidth of maximal \SI{0.5}{\kilo\hertz} based on the second moment for a Gaussian line shape \cite{Abragam1961,SlichterBook}.
We also performed hyper-polarised Hahn spin echo measurements via phase cycling to determine the spin-spin relaxation time, but we could not record any signal.
The inset in Fig. \ref{spectra} shows an average solid/dipolar echo signal of \num{32} scans in a phase cycling experiment in the time domain with \mbox{$\tau=$\SI{500}{\micro\second}} between the two $\pi/2$ pulses. As an evidence for dipolar coupling of C-13 spins a refocusing of the signal can be observed at \SI{0.5}{\milli\second} \cite{LevittBook}.\par
\begin{figure}[tb]\centering
\includegraphics[width=\linewidth]{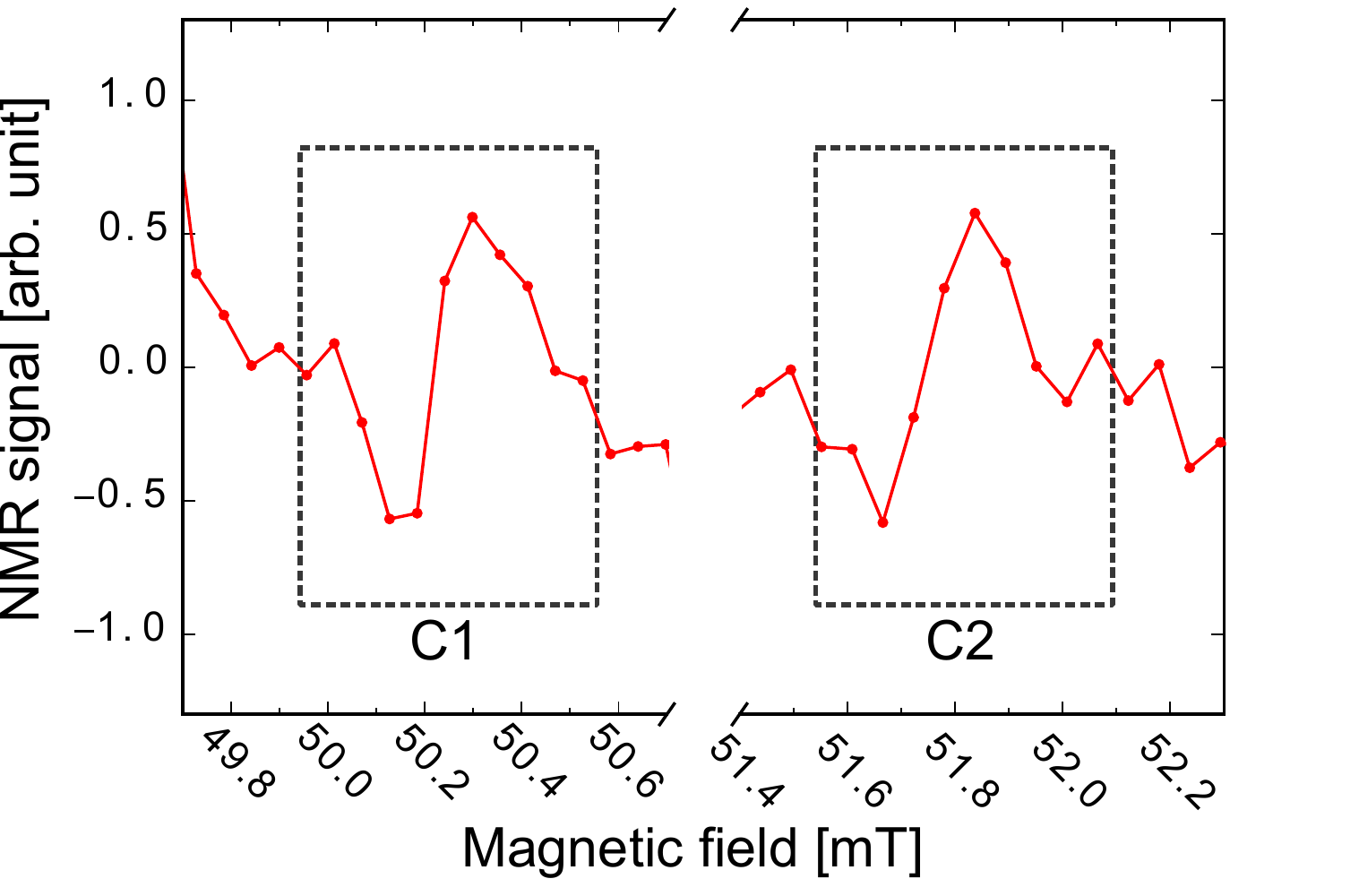}
\caption{Detailed view on smaller hyperpolarising effects in region C1 and C2 of Fig. \ref{fieldepFull}. Average of two measurements per field.}
\label{fieldepSplit}
\end{figure}
The field-dependence of the signal displays a non-trivial behaviour of pronounced peaks and sign changes (Fig. \ref{fieldepFull}). We subdivide these peaks in three classes A, B and C. The peaks of A and B can be understood if one takes into account coupling and therefore cross relaxation between the NV-carbon-13 systems and the P1-N14 systems. The energy matching between these two systems at a certain magnetic field due to each Zeeman splitting leads to a resonant coupling and consequently a cross relaxation (Hartmann-Hahn condition)\cite{HartmannHahn1962}. A P1 centre experiences a hyperfine coupling to its intrinsic \textsuperscript{14}N nuclei spin ($I$=1). Due to a Jahn-Teller distortion towards one neigboring carbon atom this coupling depends on the direction of the applied magnetic field $B$ \cite{Smith1959}. Under our experimental conditions P1 centres can be divided in two groups. One is aligned in $B$ direction and the other is one oriented under an angle of about \SI{71}{\degree} relative to the magnetic field. The latter one displays a weaker hyperfine coupling to their \textsuperscript{14}N nuclei. The smaller peaks (B) can be associated with coupling between NV centres and P1 centres both aligned parallel to the applied field where the electron spin polarisation of the NV centres is most efficient.
The case of coupling between an aligned NV centre with an unaligned P1 centre is three times more likely due to the three equivalent orientations in the diamond lattice. The peaks of the group A correspond to such a coupling and are therefore roughly three times higher than that ones of group B.
A closer look at the data represented in Fig. \ref{fieldepFull} suggests a similar but much weaker signal (C) between the peaks of highest intensity.
To support this we measured this field region twice under the same experimental conditions to enhance the signal to noise ratio (Fig. \ref{fieldepSplit}). The origin of these peaks is unclear. It could result from other spin species in the sample or \textsuperscript{15}N ($I=1/2$) associated P1 centres. The latter case would in principal explain the position and shape of this peaks but not its intensity which should be about \num{270} times weaker than the peaks referring to the \textsuperscript{14}N-P1 centres due to the natural abundance. Unlikely this hyperpolarisation signal could also be induced from the ESLAC. 
For detailed account and explanation see the theory and simulation section below.
Furthermore, the sharp resonances in the polarisation may suggest that just those carbon-13 nuclei, which are weakly coupled to the NV centre, can transport their spin polarisation to neighboring spins. Otherwise the NMR signal should be detected over a broader field range. In this picture a \grqq frozen\grqq core of strongly coupled and polarised spins surrounding the NV centre. These spins are commonly probed in ODMR experiment and show a spin polarisation in much broader field regions \cite{Jacques2009,Wang2013}.
Apparently, the Larmor frequencies of neighbouring spins in this region differ to much for a resonant spin exchange.
Thus the carbon nuclei in a more or less well defined shell-like region around the NV centre can pass on their spin polarisation into the bulk via the nuclear dipole network \cite{Khutsishvili1966}. Additionally it is statistically more unlikely to have a carbon-13 spin next to the NV centre than at some distance.\par
\begin{figure*}[tb]
\begin{center}
{\includegraphics[width=0.90\linewidth]{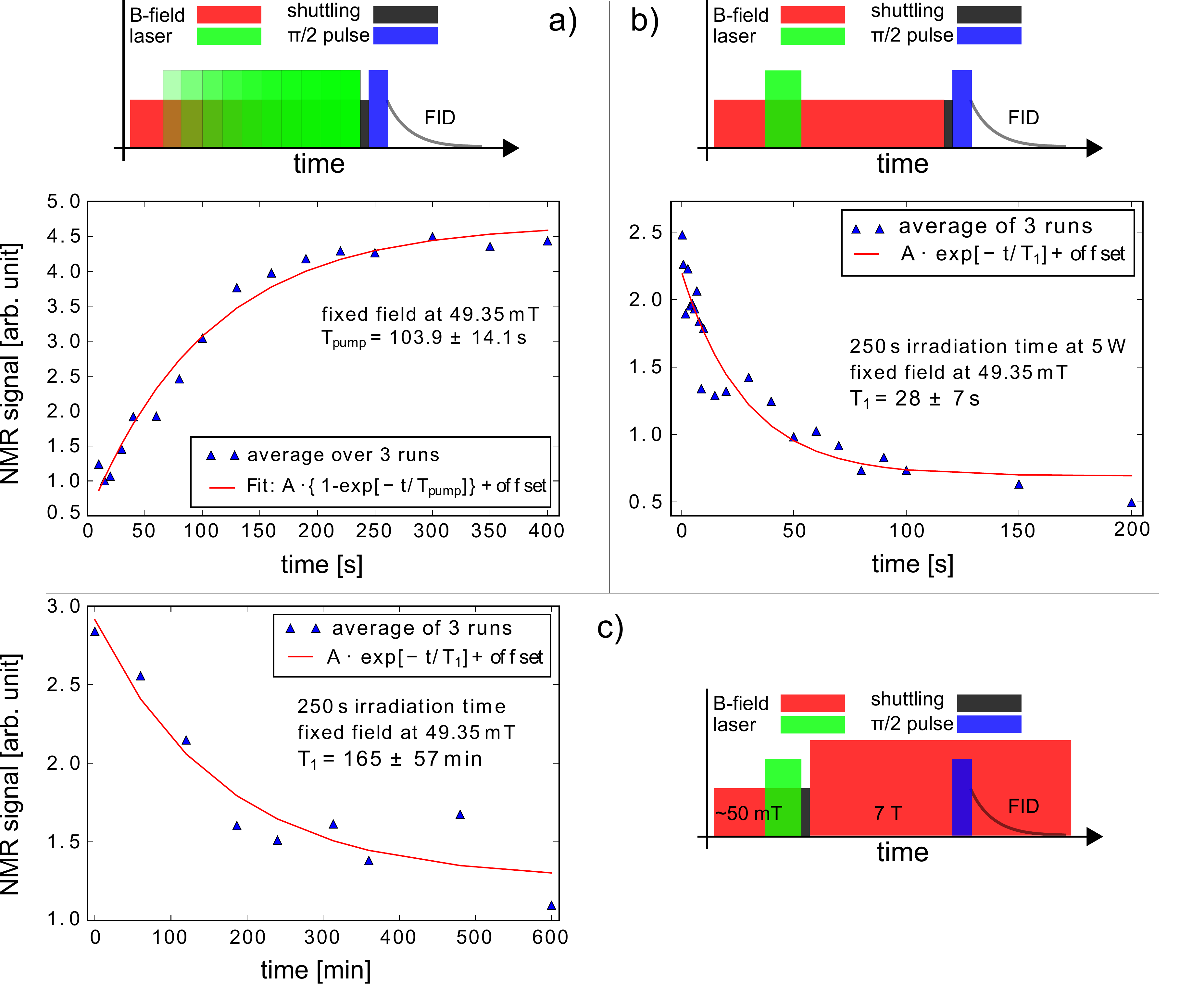}}
\caption{Time behavior of the hyper-polarisation effect depending on the irradiation time of a \SI{532}{\nano\meter} laser with \SI{5}{\watt} at \SI{49.35}{\milli\tesla} (a) and the polarisation decay in the same field after an irradiation of \SI{250}{\second} (b). The T1 time in the \SI{7}{\tesla} field is about \SI{2.75}{\hour} and were measured with the same polarisation conditions like in (b). Illustrations of the measurement procedure can be found for the corresponding plots. See Fig. \ref{timeHighVeryShort} and \ref{timeHighShort} in the appendix for high field measurements with different sets of delay times to the acquisition of the NMR signal.}
\label{timebehavior}
\end{center}
\end{figure*}

For time resolved measurement of the hyperpolarisation effect, we choose the negative peak at \SI{49.35}{\milli\tesla} in field region A1 in Fig. \ref{fieldepFull}. The experimental procedure for each experiment is illustrated schematically next to the corresponding plot in Fig. \ref{timebehavior}. To find the characteristic pumping time, the sample was irradiated in the corresponding field with constant power but different duration. Afterwards the sample was shuttled in the NMR probe and the $\pi / 2$ pulse was applied immediately (Fig. \ref{timebehavior}a). The characteristic pumping time in this experiment was determined to about $T_{\text{pump}}=$\SI{104}{\second}.
The depolarisation in the same field was measured with a pumping time of \SI{250}{\second} which is sufficient to be in the saturated region of the pumping process. The characteristic decay time back to the thermal equilibrium was determined to be $T^{\text{LF}}_1=$\SI{28}{\second} (Fig. \ref{timebehavior}b). This is about \num{3.7} times faster than the characteristic pumping time. Reasons for this can be the fact, that the NV system is frequently in the excited state during the polarisation. This is accompanied by an other electronic configuration as in the ground state, where the NV centre is in resonance with the P1 defects. Also it should be mentioned that an unstable laser output in the first seconds can cause longer pumping times.
Figure \ref{timebehavior}c) shows the exponential-like decay of the integral signal intensity over \SI{10}{\hour}. The depolarisation behaviour back to the equilibrium state in the \SI{7}{\tesla} field shows a much longer relaxation time of about $T^{\text{HF}}_1=$\SI{165}{\minute}. Due to this long time no change in the signal can be recognized within in the first \SI{300}{\second} (See Fig. \ref{timeHighVeryShort} in appendix) and even an observation time over \SI{200}{\minute} identify just a slow decrease of the signal (see Fig. \ref{timeHighShort} in appendix). 
The extreme long T1 time might be a manifestation of the wide off-resonant Zeeman splittings of the two defect systems and may be an advantage for future developments and novel applications. 

\subsection{Theory and Simulation}
\begin{figure}[bt]
\centering
\begin{overpic}[width=\linewidth]{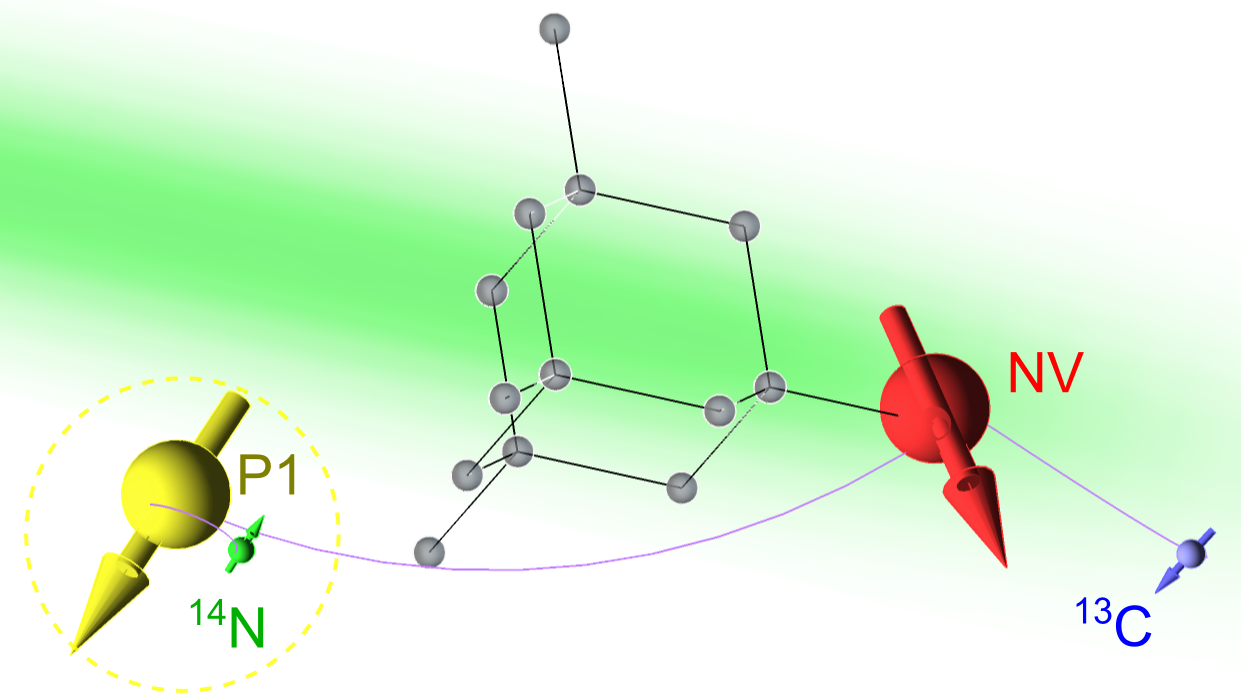}
\end{overpic}
\caption{Sketch of the simulated four spin system consisting of a P1 centre electron spin coupled to its nitrogen spin and a NV centre interacting with a carbon-13 spin.}
\label{4spinsystem}
\end{figure}

\begin{figure}[bt]
\centering
\begin{overpic}[width=\linewidth]{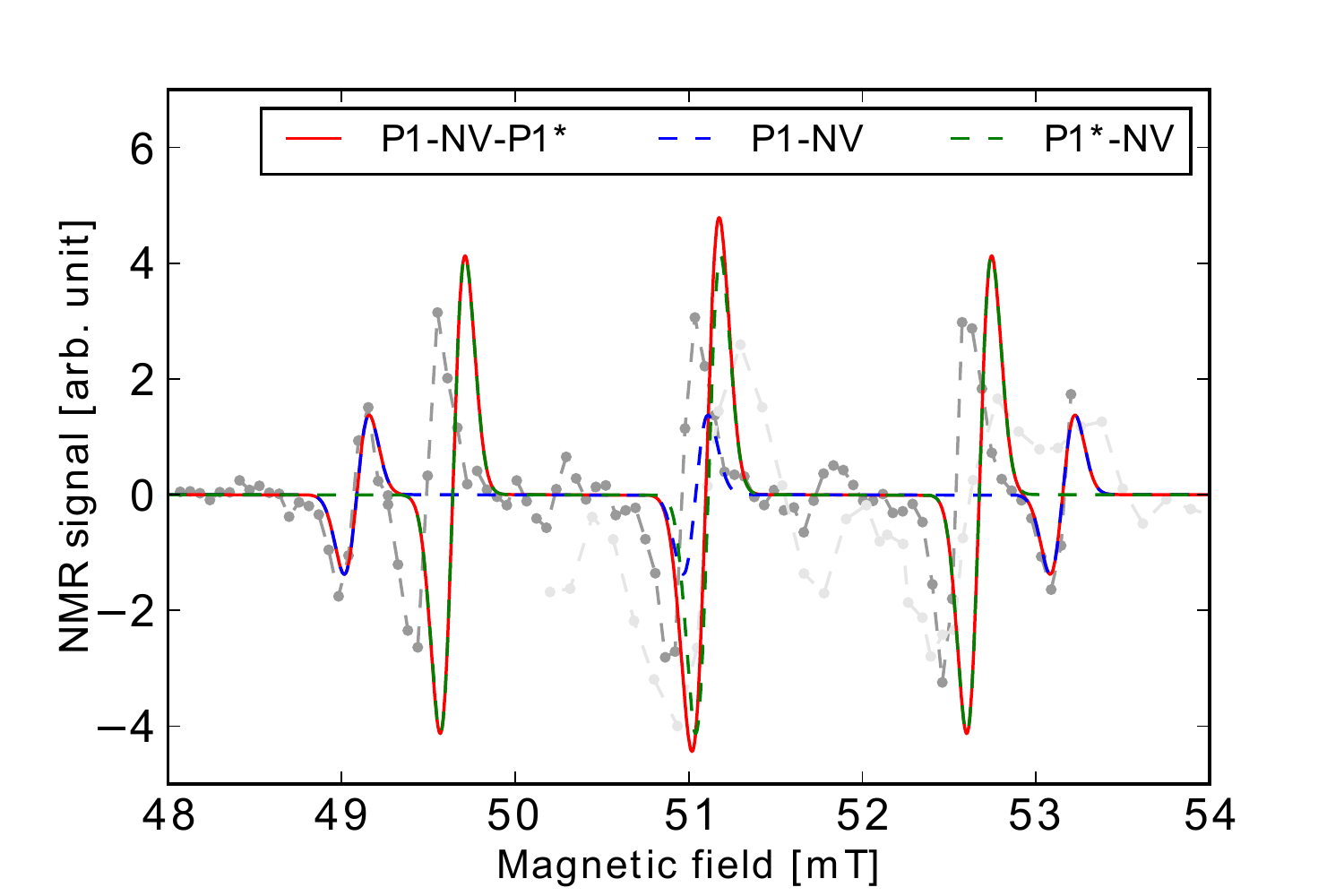}
\put(40,50){$\ket{\uparrow}$}
\put(40,90){$\ket{\downarrow}$}
\end{overpic}
\caption{Simulation of the polarisation of carbon-13 nuclei via cross relaxation among the NV-13C complex and the P1-14N system. The labels in the legend denotes the coupling of NV centres to aligned (P1) and non-aligned (P1*) substitutional nitrogen defects as well as the superposition of both (red). The data from Fig. \ref{fieldepFull} is highlighted in grey.}
\label{SimPol}
\end{figure}

We assume two quantum objects one consisting of a NV centre in its optical ground state with the electron spin $S^{\text{NV}}$ and a nearby spin $I^{\text{C}}$ of a carbon-13 nucleus. The other one is composed of the electron spin $S^{\text{P1}}$ of a P1 centre and the hyperfine coupled intrinsic nitrogen spin $I^{\text{N}}$. Here we consider the N-14 isotope ($I^{\text{N}}=1$) due to a natural abundance of \SI{99.63}{\percent}. In this model we neglect the interaction of the intrinsic nitrogen spin of the NV centre which is weak in the ground state. The situation is illustrated in Fig. \ref{4spinsystem}.
This leads to the following two Hamiltonians,
\begin{align*}
H^{\text{NV}} &= D \, \left(S_{\text{z}}^{NV}\right)^2 + \gamma \, B \, S_z^{\text{NV}} +  A^{\text{13C}} \, \vec{S}^{\text{NV}} \cdot \vec{I}^{\text{C}}\\
H^{\text{P1}}_{\text{A,B}}  &= \gamma \, B \, S_z^{\text{P1}} +  A^{\text{A,B}}  \,\vec{S}^{\text{P1}} \cdot \vec{I}^{\text{N}}  \text{,}
\end{align*}
where $A^{\text{13C}}$ denotes the isotropic interaction parameter for the NV and \textsuperscript{13}C spin, $A^{\text{A,B}}$ the hyperfine coupling parameter for the aligned (A) and the non-aligned P1 centres (B) along the z-axis and $\gamma$ the gyromagnetic ratio of an electron (\SI{28.03}{\mega\hertz \per \milli\tesla}) The applied external magnetic field $B$ is oriented parallel to the z-axis. The zero field splitting $D$ in the NV optical ground state is \SI{2870}{\mega\hertz}.
Additionally, we assume in this model a dipole-dipole coupling of the NV coupled carbon-13 spin to an initially unpolarised C-13 spin reservoir, which is not directly considered in the Hamiltonian but ensures the distribution of spin polarisation over the hole diamond lattice.
For our simulations we choose a relatively weak coupling of the NV centre to the carbon spin of $A^{\text{13C}}=\SI{2}{\mega\hertz}$ consistent with the narrow polarisation signal in the field dependent measurements. But it should be mentioned, that simulations, where the coupling does not exceed values of \SI{20}{\mega\hertz} display a very similar polarisation pattern. For higher values the energy level is further shifted which produces unobserved peaks and peak shapes.
The values for the hyperfine coupling are taken from the literature as $A^{\text{A}}=\SI{85}{\mega\hertz}$ and $A^{\text{B}}=\SI{114}{\mega\hertz}$ \cite{Holliday1989,Smith1959}.
By solving the Hamilton eigenvalue problem the energy levels of both systems are calculated. In the case of the P1 complex we distinguish between the two possible orientations in the crystal lattice and its relative abundance (1:3).
The nitrogen content in the sample results in an average distance of nitrogen associated defects of about \SI{10}{\nano\meter}. This corresponds to a dipolar coupling strength of \SI{200}{\kilo\hertz} -- or for a distance of \SI{5}{\nano\meter} even \SI{2.5}{\mega\hertz} are found.
A resonant energy transfer between both quantum systems achieved within an energy difference of \SI{\pm 0.5}{\mega\hertz} in the presented simulation (Fig. \ref{SimPol}), being characteristic for the coupling strength among the defects. For each initial and final state of the possible transitions the expectation value $I^{\text{C}}$ is stored to determine the preferable spin polarisation. In this simple algorithm the differences of $I_{\text{z}}$ reflects the spin flip probability of the carbon-13 nucleus.
Now each polarised carbon-13 nucleus acts as source of either positive or negative spin polarisation for dipole coupled neighbouring spins with similar Larmor frequencies.
The resolution of the field sweep is modelled as a convolution with a Gaussian with a full width at half maximum of \SI{0.1}{\milli\tesla}.
With this approach the shape of the marked regions in the experimental data in Fig. \ref{fieldepFull} can be well reproduced.
In this model it is not obvious which physical parameter determines the recorded line width of the polarisation peaks in Fig. \ref{fieldepFull}. Either both the electron interaction NV-P1 and the NV-C-13 coupling are weak or the interaction to the P1 system is weak (narrow resonant energy transfer) and the coupling of the NV centre to the carbon-13 spin is strong or vice versa.
Nevertheless, the concept of a coupling scenario in the NV optical ground state is reasonable because of the rather short life time of several nano seconds in  the excited state \cite{NeumannExcState2009}.
This short life time corresponds to a short interaction time with very slow time evolution of the state vectors, which are responsible for the polarisation effect in terms of hyperfine coupling and cross polarisation. This only allows efficient interaction with strongly coupled spins. In contrast this is not the case in the ground state where also weak coupling could contribute to a polarisation effect.

\subsection{Conclusion and Outlook}
In the present contribution we highlight a magnetic field dependent nuclear polarisation around \SI{50}{\milli\tesla} measured by NMR.
Our data suggest that the observed effect can not be allocated to the hyperfine coupling in ESLAC but is much more likely related to a cross polarisation between a NV carbon-13 system and a nearby P1 centre. 
The dominant polarisation pattern can be explained by and attributed to cross relaxation between dipolar coupled NV and \textsuperscript{14}N-P1 centres in good agreement with the theoretical simulation. 
Two minor features in the data are observed whose origin is still unclear at present but may be associated with \textsuperscript{15}N-P1 centres.
If this conclusion is true it would suggest that an interaction with a P1 centre generates an efficient bulk polarisation even at relatively low concentrations. 
In this picture the P1 defect acts as mediator for the nuclear polarisation effect much more efficient than the ESLAC and should be exploited further in future hyperpolarisation applications.\par
In order to investigate this effect in more detail the defect distance has to be modified, which can be accomplished by samples with different nitrogen content. A fluence variation of the electron irradiation would change the NV-P1 ratio and therefore also the average distance between those defect types.
One cannot rule out the possibility that with shrinking distance the interaction becomes so strong that the process of spin flips has to be described by quantum state overlaps and time depending Schrödinger equation. For the current experiments this was not necessary.
Furthermore it becomes apparent that maybe a selective coupling to C-13 or other spins with a minimum distance to the NV centre is necessary for an effective spin diffusion into the surrounding bulk.

\phantomsection
\section*{Acknowledgments} 
\addcontentsline{toc}{section}{Acknowledgments} 
This work was supported by the VolkswagenStiftung. We thank Dr. W. Knolle from the Leibniz Institute of Surface Modification (IOM) for helpful discussions and valuable assistance during the high energy electron irradiation.


\phantomsection
\newcommand*{\doi}[1]{\href{http://dx.doi.org/#1}{doi:#1}}
\bibliographystyle{plainnat}

\bibliography{CrossRelaxHyperpol}

\newpage
\begin{appendices} 

\begin{figure}[ht]\centering
\includegraphics[width=\linewidth]{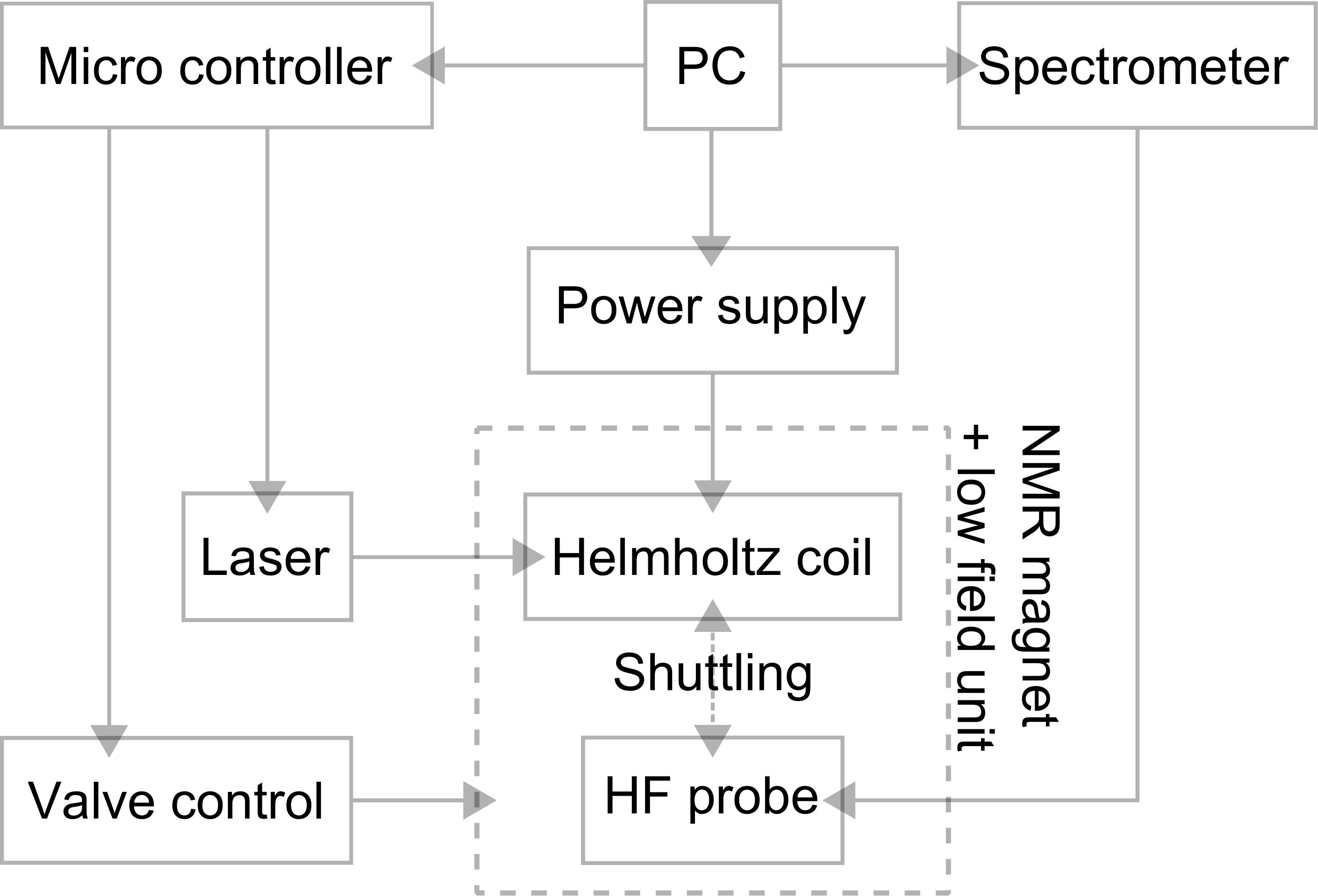} 
\caption{Block diagram of the experimental control.}
\label{blockdia}
\end{figure}
\begin{figure}[ht]\centering
\includegraphics[width=0.8\linewidth]{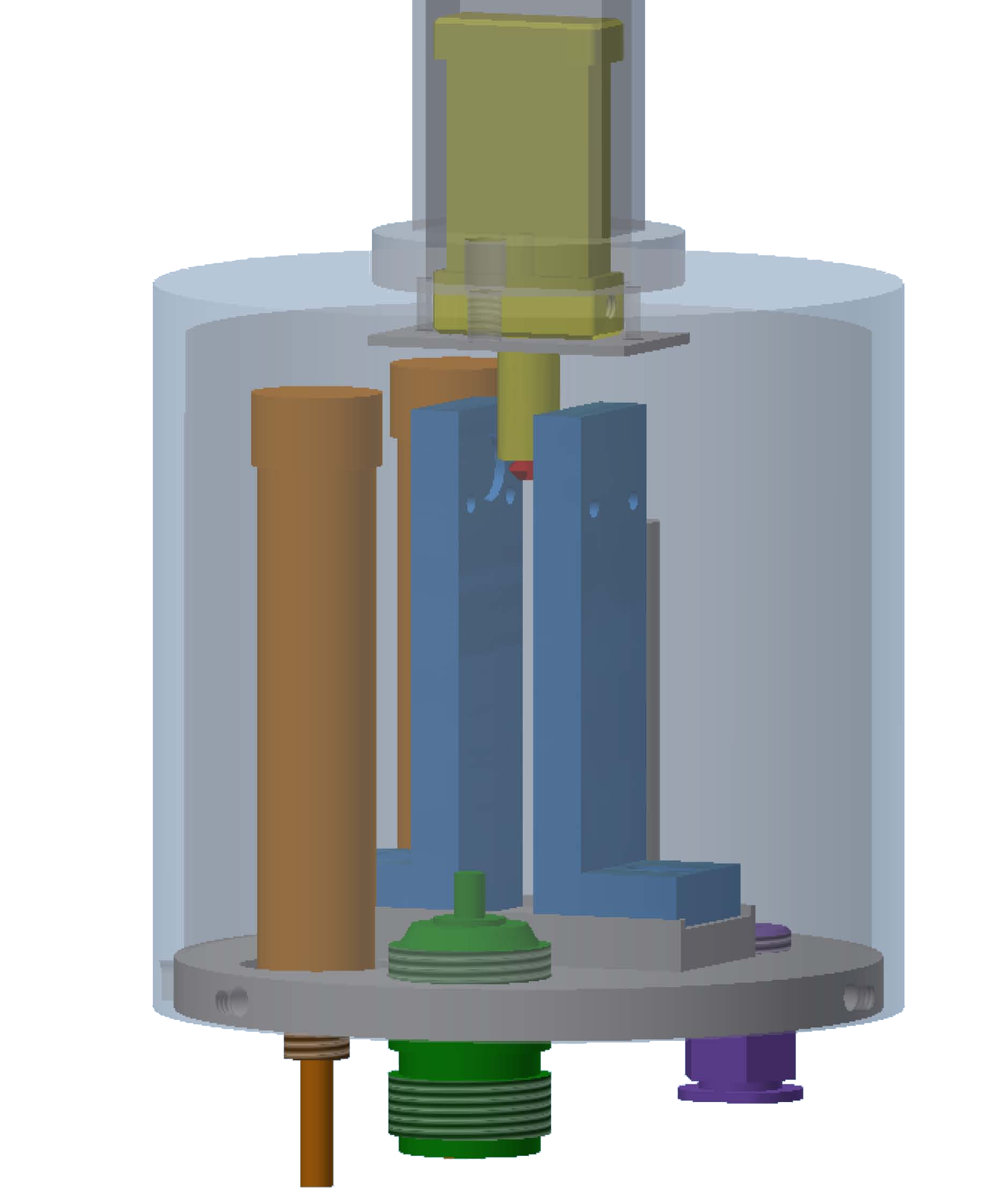} 
\caption{Sketch of the NMR probe with sample (red) mounted on the shuttle (yellow) and positioned between the ceramic HF coils holder (blue). Capacitors for tuning and matching shown in orange. In the bottom of the probe a connector for electrical access (green) and a pneumatic port shuttling is integrated.}
\label{nmrProbe}
\end{figure}
\begin{figure}[ht]\centering
\includegraphics[width=\linewidth]{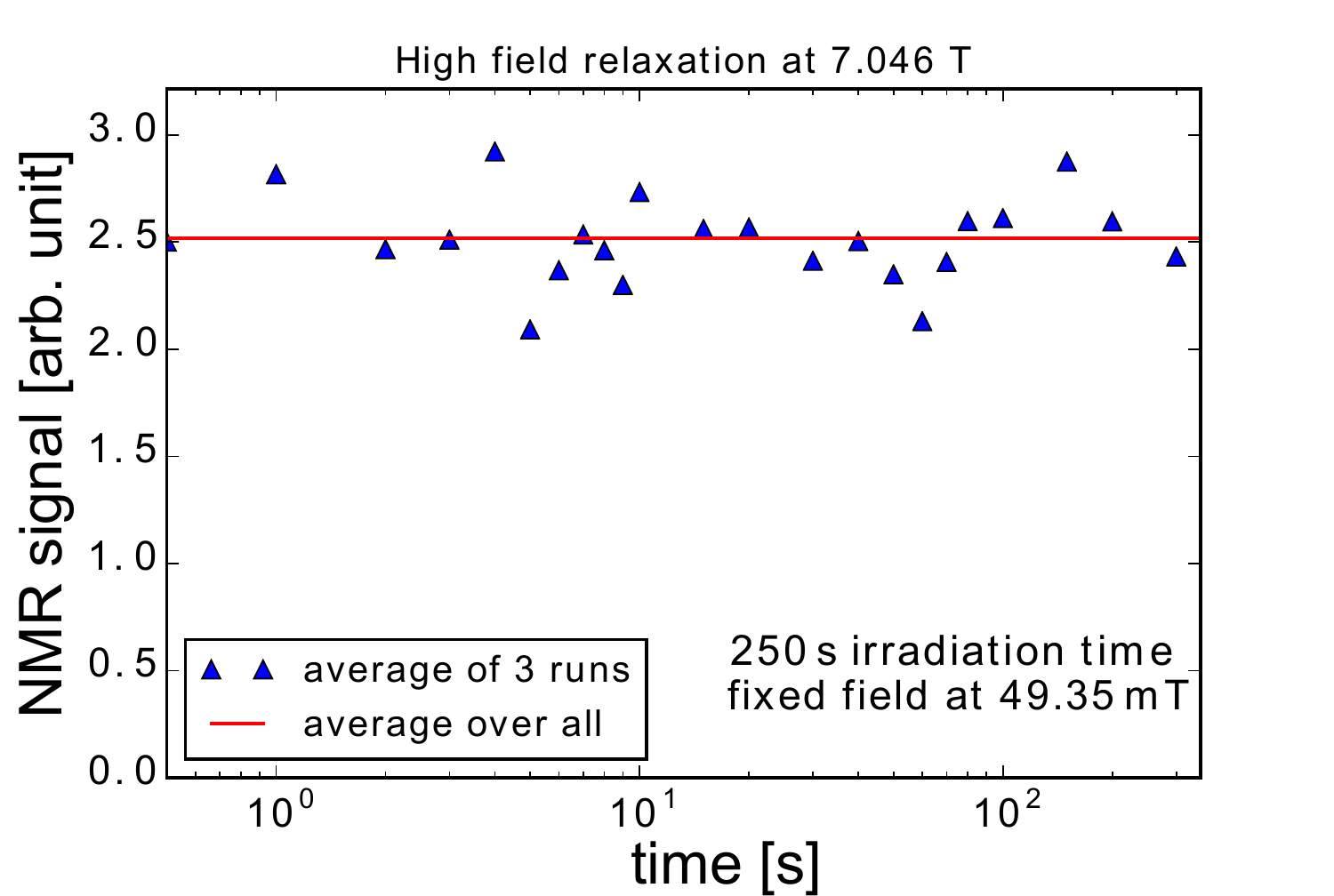}
\caption{Time behaviour of the hyper-polarisation effect at the \SI{7}{\tesla} field within the first \SI{300}{\second} after the shuttling.}
\label{timeHighVeryShort}
\end{figure}
\begin{figure}[ht]\centering
\includegraphics[width=\linewidth]{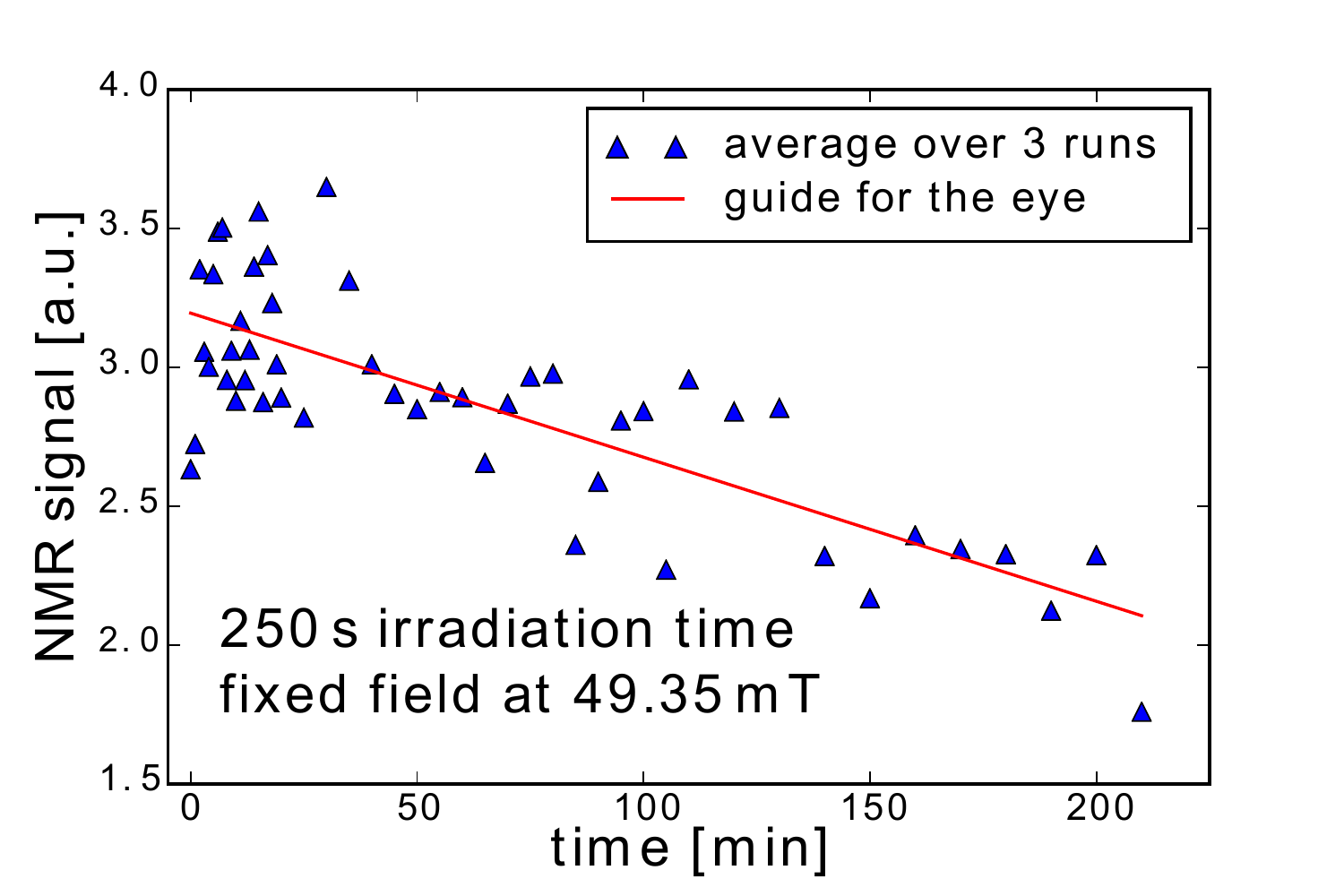} 
\caption{Time behaviour of the hyper-polarisation effect at the \SI{7}{\tesla} field within the first \SI{200}{\minute} after shuttling.}
\label{timeHighShort}
\end{figure}

\end{appendices}

\end{document}